# OpenMUSTANC (MUltiple Scattering Theory At Nanoplasmonic Cavities): A MATLAB toolbox for the simulation of Plasmonic Sphere Aggregates

Xin Zheng[1], Christos Mystilidis[2], Christos Tserkezis[2], Guy A. E. Vandenbosch[1], Xuezhi Zheng[3]

[1] WaveCoRE Division, Department of Electrical Engineering, KU Leuven, 3001, Leuven, Belgium

[2] POLIMA–Center for Polariton-Driven Light-Matter Interactions, University of Southern Denmark, 5230 Odense, Denmark

[3] Key Laboratory of Radar Imaging and Microwave Photonics, Ministry of Education, Nanjing University of Aeronautics and Astronautics, Nanjing, Jiangsu 211106, China

**Abstract:**

Mesoscopic physical models, including the Hydrodynamic Drude Model (HDM), the Generalized Nonlocal Optical Response (GNOR) Model, and the Surface Response Model (SRM), have been proposed to investigate nonlocal effects in nanometric structures. The combination of classical electromagnetism with these mesoscopic material models calls for new computational electromagnetic (CEM) algorithms, or update of conventional ones, in what is termed computational mesoscopic electromagnetics (CMEM). In this work, we present a MATLAB toolbox for the simulation of multiple spherical interfaces with arbitrary relative positions, with the incorporation of the aforementioned mesoscopic models. The method exploits vector spherical wave functions to properly express electric and magnetic fields, an **S** matrix formulation for the efficient treatment of incident and scattered fields at spherical interfaces, and a translation matrix to deal with propagating waves with different expansion centers. Excitation sources can be chosen among arbitrarily polarized plane waves, dipoles and electron beams. For the post-processing part, the calculation of cross sections and far/near-field mapping; fluorescence enhancement, Purcell factor and quantum yield; and cathodoluminescence and electron energy-loss probability is implemented. The toolbox is built in a modular manner, and each part (routine) has its own important functionality. This paper provides a full explanation of the proposed highly efficient and general toolbox, and a detailed guideline for researchers in the nanoplasmonics community.

Program title: A MATLAB toolbox for the simulation of Plasmonic Sphere Aggregates
Licensing provisions: GPL-3.0
Programming language: MATLAB R2023a
External routines/libraries: Gmsh (mesh generation)
Nature of problem: This program addresses the semi-analytical simulation of nonlocal optical response in plasmonic sphere aggregates. When nanostructures enter the sub-nanometer regime, quantum phenomena such as frequency shifts due to screening or electron spill-out, and spectrum broadening owing to Landau damping become dominant. This toolbox is designed to efficiently predict the nonlocal optical response of complex, multi-layered, or clustered plasmonic sphere aggregates with arbitrary spatial configurations, whose material can be described by the main mesoscopic models, including the Hydrodynamic Drude Model (HDM), the Generalized Nonlocal Optical Response (GNOR) model, and the Surface Response Model (SRM). The excitation can be chosen between arbitrarily polarized plane waves, point dipoles, or electron beams.
Solution method: This toolbox uses the S-Matrix formalism with vector spherical wave expansions, and the translational addition theorem links fields expanded with respect to different centers. The main equation behind the algorithm is solved linearly by using Generalized Minimal Residual (GMRES) iteration or by direct inversion. Dedicated post-processing modules are integrated to calculate optical cross sections, electron energy-loss (EEL) and cathodoluminescence (CL) probabilities, as well as fluorescence enhancement.
Restrictions: A homogeneous and lossless background; non-overlapping nanospheres; simulation accuracy depends on $l_{\max}$ truncation; non-penetrating electron trajectories.
Typical running time: On AMD Ryzen 7950X 16-Core 32-thread Processor, 128GB 6000MHz ddr5 RAM, Windows 11 Pro, example 1 (concentric core-shell, $l_{max} = 12$, 61 wavelengths): 412.005 s; example 2 (concentric core-shell, $l_{max} = 10$, 401 wavelengths): 592.440 s; example 3 (helix structure – gap 0.3 nm, $l_{max} = 22$, 101 wavelengths): 39298.944 s.
RAM usage: Example 1 peak memory ~ 11.93 GB; Example 2 peak memory ~ 2.175 GB; Example 3 peak memory ~ 5.40 GB.

## 1 Introduction:

With the advancement of nanofabrication technology, metallic nanoparticles (NPs) of a few nanometers [1] and gap sizes of a few angstroms become realizable [2,3], opening a window to investigate the interaction of light, quantum emitters, and electron beams with metallic NPs , enabling both novel applications in light–matter interactions and deeper understanding of the underlying physical mechanisms. The local response approximation (LRA), the workhorse of most investigations in nanoplasmonics in the last decades, can be tremendously successful in both the microwave band and the optical regime; however, for shrinking dimensions comparable to the Fermi wavelength in metals, new physical models describing the quantum dynamics of free electrons need be implemented. in order to explain deviations from classical predictions, in both experiment and theory, and manifesting as frequency shifts [1,4,5], quantum tunneling [6,7] and increased damping, and the quenching of extreme field enhancement in sub-nanometer gaps [8,9]. Standard and reliable nonlocal response models include the Hydrodynamic Drude Model (HDM) [10], the Generalized Nonlocal Optical Response (GNOR) model [11,12] and the Surface Response Model (SRM) [13,14]. HDM incorporates the Pauli exclusion principle by considering quantum pressure which translates into a free-electron convective flow, while the GNOR additionally takes surface-enhanced damping through free-electron diffusion. Unlike the HDM and the GNOR, which add additional terms in the constitutive relations, the SRM focuses on the field behaviors at the boundary and summarizes all the mesoscopic physical details by introducing Feibelman parameters for the centroids of induced charge and current.

Current optical simulation frameworks suffer from a pronounced scale separation. Ab-initio approaches like time-dependent density functional theory (TDDFT) are restricted to small atomic clusters or highly symmetrical nanostructures, while standard commercial solvers, such as Lumerical (based on the Finite Difference Time Domain method – FDTD) or FEKO (based on the Method of Moments algorithm - MoM), assume the classical local response model. This leaves a critical theoretical and computational gap in the mesoscopic regime that bridges microscopic electronic behaviors and macroscopic material responses. To fill in this gap, a nascent branch termed computational mesoscopic electromagnetics (CMEMs) [15,16] has appeared, where the aforementioned semi-classical models have already been incorporated in conventional CEM algorithms to investigate light-matter interaction. FDTD methods have been presented from the very first steps of CMEM and have recently multiplied, focusing around variations of hydrodynamic models [17–20]. Finite-Element Methods (FEM), and especially Hybridizable Discontinuous Galerkin (HDG) variations have also gained interest and have been deployed for hydrodynamic models of increasing sophistication [21–24]. From the side of Integral Equation Methods, both Boundary Element Methods (BEM) [25–28] and Volume Integral ones [29,30] have appeared and have explicitly targeted hydrodynamic models, but more recently also the SRM [31]. The Discrete Sources Method (DSM) has been extended to simulate efficiently nanoplasmonic architectures both within HDM and GNOR, and within the SRM [32,33]. Efficient, semi-analytical techniques for a limited set of architectures have also been developed, revolving around transfer or cascaded scattering matrices for both hydrodynamic and SRM approaches [34,35].

Naturally, the motivation behind the development of such wealth of solution strategies is to offer ad hoc tools to members of the nanoplasmonics community—theorists and experimentalists alike. This requires, however, a concentrated effort towards integrating said recipes in approachable and (as much as possible) easy to use software, that is additionally free and open. Though certain contributions exist in this aspect in CMEM, the variety of numerical techniques being presented has not been yet translated to available software, limiting their widespread adoption, verification, and scaling. Important exceptions are

- *Moosh* [36] (recently succeeded by *PyMoosh* [37])
  It offers a scattering-matrix-based platform for the probing of (potentially nonlocal within HDM) multilayered substrates, probed by plane waves, Gaussian beams, and quantum emitters.
- *MNPBEM* [38]
  It implements a mixed-potential BEM, able to tackle arbitrarily shaped metallic nanoscatterers, potentially including substrates [39], probed by light, quantum emitters, and electrons [40]. Nonlocality is included through the Local Analogue Model [41].
- *NANOBEM* [42]
  It relies on an extension of the Stratton-Chu integral equations in order to simulate metallic and dielectric structures, excited by plane waves or dipoles and described by the LRA or the SRM [31].
- *OpenSANS* [43]
  The predecessor of the current software, based on scattering matrices and able to analyze multilayered planes, cylinders, and spheres within the LRA or HDM and excited by plane waves.

Very popular among practitioners are COMSOL implementations of such semi-classical models, with numerous studies reporting results [8,44–47] and some sharing the codes [12,48].

In this work, we present an **S** matrix-based simulation toolbox for metallic sphere aggregates. The toolbox is developed based on a rigorous semi-analytical framework leveraging the expansion of vector spherical wave functions, the **S** matrix and translational addition theorem. Notably, the toolbox offers three core capabilities: first, although the application scenario is limited to an infinite homogeneous medium as the background, it can deal with multiple spherical interfaces with arbitrary positions, which means eccentric nanospheres with multiple cores can coexist with concentric nanospheres in one single simulation, extending the application ranges; second, it incorporates three different excitation methods, plane waves, point dipoles simulating quantum emitters, and electron beams; third, regarding the post-processing part, this toolbox can be used to calculate optical cross sections, Purcell factor [49], quantum yield [50], fluorescence enhancement [51,52], CL and EEL probabilities [53–56], serving as a complete platform to investigate various physical mechanisms, such as the surface-enhanced Raman scattering [57,58], the chiral properties of helical structures or nanolenses [59,60], and the nonlocal optical response of plasmonic nanostructures interacting with quantum emitters [15,61–64]. The proposed toolbox can emerge as a powerful tool in the study of mesoscopic electrodynamics and, at the same time, can serve as a critical benchmark and reference for future advancements in numerical techniques in nanoplasmonics.

The toolbox is implemented in MATLAB R2023a, employing a modular, functional design centered on vectorized computations and structure-based variables. It supports three physical models (LRA, HDM, SRM) and three excitation types (plane waves, dipoles, electron beams). Meshes generated with *Gmsh* [65] are used for field mapping and cross section calculations. This paper presents a detailed and instructive description of the simulation toolbox, and the functions, inputs, and outputs of each demo are well commented on. In Section II, the theory behind the solver is generally explained and more details can be found in our most recent work [66]. In Section III, an overview of the proposed toolbox is provided, including the setup procedures for inputs, simulation structures, excitations, solving, and post-processing parts. In Section IV, three examples are given. For a quantitative check, we consider a concentric shell structure with a point dipole placed close to the nanostructure, functioning as a fluorescent molecule. In the first stage, the nanostructure is excited by plane waves and the resulting enhanced electric field excites the molecule; while in the second stage, the dipole radiates, where the Purcell effect induced by the metallic nanoparticle alters both the radiative and non-radiative power of the molecule. The obtained fluorescence enhancement, quantum yield and Purcell factor are compared with analytical results, showing perfect agreement. Second, a shell nanostructure is excited by electron beams, the detected EEL and CL probability agree well with the analytical results, proving the toolbox's accuracy. For a physical check, excitations of plane waves with different polarizations are performed on a helix structure, to investigate the quantum chiral properties.

## 2 Theoretical Background:

While the algorithm behind the proposed toolbox has been discussed in detail in our previous work [66], a brief overview is provided here for completeness. The toolbox is designed for multiple nonoverlapping spherical interfaces belonging to the same or multiple spheres (see Fig. 1), whose inner region and/or outer region can be filled with (nonclassical) metals or dielectric materials. Under a known external excitation, the primary objective of the algorithm is to calculate the generated scattered fields due to light-matter interactions.

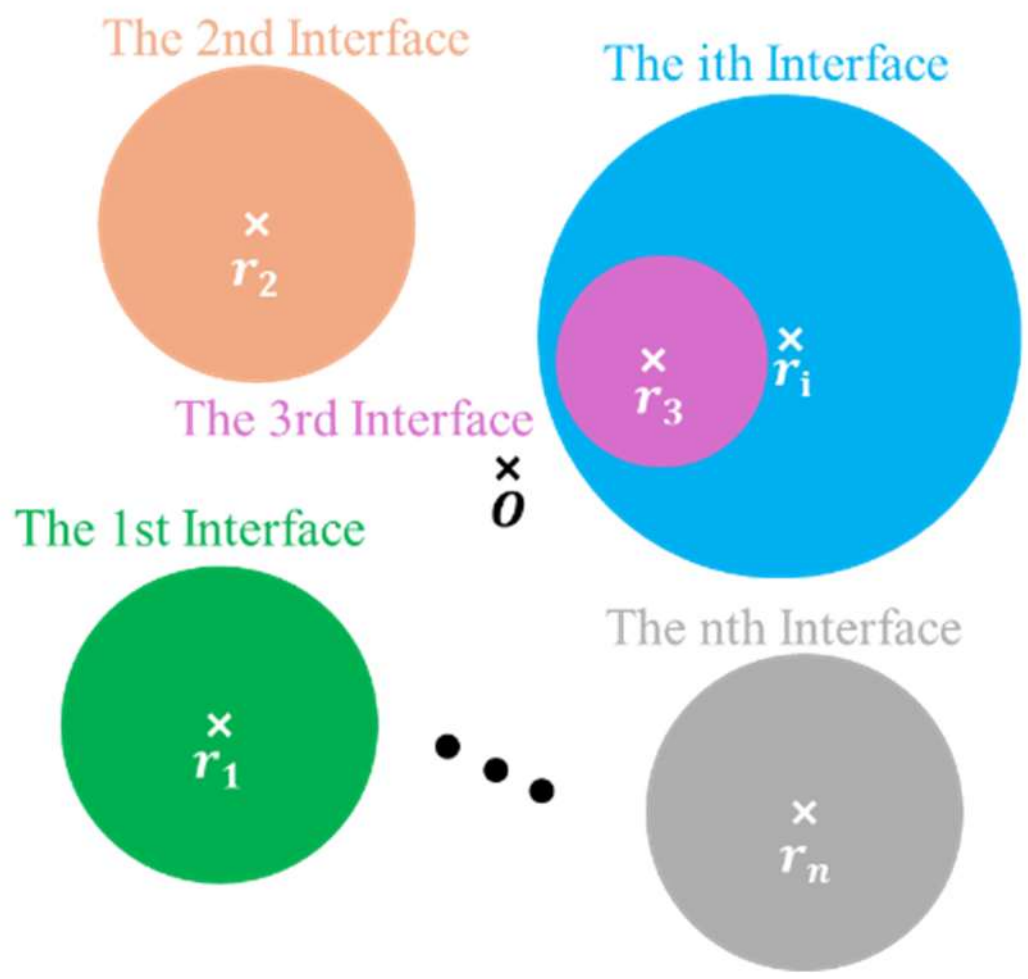

Fig. 1. Schematic illustration of an aggregate system of $N$ spherical interfaces.

## 2.1 Maxwell Equations

The main goal of CEM is to propose efficient and effective algorithms to solve Maxwell equations. The macroscopic equations in a source-free space are listed as below,

$$\begin{aligned} &\nabla \cdot \mathbf{D}(\mathbf{r},\omega) = 0, \\ &\nabla \cdot \mathbf{B}(\mathbf{r},\omega) = 0, \\ &\nabla \times \mathbf{E}(\mathbf{r},\omega) = i\omega \mathbf{B}(\mathbf{r},\omega), \\ &\nabla \times \mathbf{H}(\mathbf{r},\omega) = -i\omega \mathbf{D}(\mathbf{r},\omega). \end{aligned} \tag{1}$$

Here, $\omega$ is the angular frequency, and we adopt an $e^{-i\omega t}$ time dependence throughout this paper. $\mathbf{r}$ is a random point in space, $\mathbf{E}$ and $\mathbf{H}$ are the electric field and magnetic field, respectively. $\mathbf{B}$ is the magnetic flux density, and $\mathbf{D}$ is the electric displacement field. The constitutive relations between these physical parameters are,

$$\begin{aligned} &\mathbf{B}(\mathbf{r},\omega) = \mu_0 \mu_r \mathbf{H}(\mathbf{r},\omega), \\ &\mathbf{D}(\mathbf{r},\omega) = \varepsilon_0 \mathbf{E}(\mathbf{r},\omega) + \mathbf{P}(\mathbf{r},\omega), \end{aligned} \tag{2}$$

where $\varepsilon_0$ and $\mu_0$ are the vacuum permittivity and permeability. Since natural materials exhibit negligible magnetic response at optical frequencies, the relative magnetic permeability of all media is assumed to be unity ($\mu_r = 1$) [67]. $\mathbf{P}$ is the total polarization current, composed of the polarization current of free electrons and bound electrons.

## 2.2 Mesoscopic Physical Models and Boundary Conditions

The electromagnetic (EM) response of the materials used in this toolbox is described by three different models, the LRA, HDM and SRM. In the HDM, the classical constitutive relation is replaced by a partial differential equation, which constitutes the key equation of nonlocal electrodynamics,

$$\xi^2(\omega)\nabla[\nabla \cdot \mathbf{J}(\mathbf{r})] + \mathbf{J}(\mathbf{r}) = \sigma(\omega)\mathbf{E}(\mathbf{r}). \tag{3}$$

In Eq. (3), **J(r)** is the current density, $\sigma(\omega)$ is the frequency-dependent free electron Drude conductivity, where $\sigma(\omega) = i\varepsilon_0\omega_p^2/(\omega + i\gamma)$ with $\omega_p$ being the plasma frequency and $\gamma$ the damping rate. $\xi$ is the dispersive strength of the quantum correction, which is approximately the distance that an electron would move by convection during the time of an optical cycle. For noble metals, this length scale is typically on the order of 0.1 nm at plasmonic resonances (and at most 1 nm in the lower frequency regime) [15,47], and is given by,

$$\xi^2(\omega) = \frac{\beta^2}{\omega(\omega + i\gamma)}, \beta^2 = \frac{3}{5}v_F^2. \tag{4}$$

In the above, $\beta$ is a parameter related to the quantum pressure, which reflects the influence of Pauli exclusion principle, while $v_F$ is the Fermi velocity. In the examples presented in Section 4, simulated by the proposed toolbox, HDM is used. If diffusive effects are considered as well, the HDM extends to the GNOR model, and the dispersive length becomes,

$$\xi^2(\omega) = \frac{\beta^2}{\omega(\omega + i\gamma)} + \frac{\mathscr{D}}{i\omega}. \tag{5}$$

where, $\mathscr{D}$ is the diffusive constant associated with the diffusive transport of the electrons. We clarify that diffusive kinetics is an efficient manner to include surface-enabled Landau damping, as shown in [68].

Since the medium can be treated as either local or nonlocal, the boundary is divided into four categories, namely a local-local boundary, a local-nonlocal boundary, a nonlocal-local boundary and a nonlocal-nonlocal boundary. The boundary conditions needed to be satisfied for each kind are different. In the case of a local-nonlocal boundary, if using the HDM or the GNOR, an additional boundary condition regarding the current density needs to be added, apart from the two classical boundary conditions considered in the local-local boundary, which guarantee the continuity of the tangential components of electric field and magnetic field at the interface,

$$\begin{aligned} \mathbf{n} \times \mathbf{E}_1 &= \mathbf{n} \times \mathbf{E}_2, \\ \mathbf{n} \times \mathbf{H}_1 &= \mathbf{n} \times \mathbf{H}_2. \end{aligned} \tag{6}$$

Here, $\mathbf{n}$ is a normal vector of the boundary, and subscripts 1 and 2 refer to the inner region and the outer region of an interface, with **n** directed from region 1 to region 2. For the HDM, it is assumed that all electrons are trapped in the inner side of the interface, and the polarization current is terminated at the interface, implying,

$$\mathbf{n} \cdot \mathbf{J} = 0. \tag{7}$$

We take a single interface (metal on the inner side and dielectric medium on the outer side) as an example. Then, Eq. (7) implies [69],

$$\mathbf{n} \cdot \varepsilon_{1,bd}\mathbf{E}_1 = \mathbf{n} \cdot \varepsilon_2\mathbf{E}_2. \tag{8}$$

$\varepsilon_{1,bd}$ is the bounded electron permittivity of the inner metal. $\varepsilon_2$ is the relative permittivity of the outer dielectric medium. The boundary conditions of the nonlocal-local boundary can be deduced in the same way. For the nonlocal-nonlocal boundary, the longitudinal waves are introduced on both sides, therefore, the continuity of the normal component of the free-electron current density and the continuity of charge density at the surface needs to be satisfied as well [69],

$$\mathbf{n}\cdot\varepsilon_{1,bd}\mathbf{E}_1 = \mathbf{n}\cdot\varepsilon_{2,bd}\mathbf{E}_2,$$
$$c_1\nabla\cdot\mathbf{E}_1 = c_2\nabla\cdot\mathbf{E}_2. \tag{9}$$

Here, the coefficient $c = \frac{\beta^2}{\omega_p^2}\varepsilon_{bd}$.

For the SRM, the bulk of the metals is treated by using LRA, but there exists a transition region around the boundary, where Friedel oscillations and the evanescent tail associated with quantum spill-out happen. In this model, the nonlocal effects are introduced by including quantum-corrected boundary conditions, namely [13],

$$\mathbf{E}_2^{\parallel} - \mathbf{E}_1^{\parallel} = -d_{\perp}\cdot\nabla_{\parallel}(\mathbf{E}_2^{\perp} - \mathbf{E}_1^{\perp}). \tag{10}$$

$$\mathbf{H}_2^{\parallel} - \mathbf{H}_1^{\parallel} = -i\omega d_{\parallel}\cdot\mathbf{n}\times(\mathbf{D}_2^{\parallel} - \mathbf{D}_1^{\parallel}). \tag{11}$$

Here, the $d$ parameters are Feibelman parameters and can be extracted from TDDFT or other relevant models (see [14,15,30,70]), related to the centroid of the induced charge density and the corresponding tangential current in the selvage region. The superscripts $\parallel$ and $\perp$ represent the transverse and longitudinal components (parallel with or perpendicular to the interface), respectively. In this paper, we use $d_{\parallel}$ equal to zero and $d_{\perp}$ curve-fitted by formulas from reference [14], originally pertinent to a planar half-space, and neglect spatial dispersion therein [71].

### 2.3 S Matrix

In the proposed toolbox, the electric and magnetic fields (either incident or scattered) are expanded in terms of spherical wave functions. The longitudinal wave function **L** is additionally added when using a nonlocal response model,

$$\mathbf{E}(\mathbf{r}-\mathbf{r}_c) = \sum_{lm}\mathbf{M}_{lm}(k,\mathbf{r}-\mathbf{r}_c)\cdot a_{lm} + \mathbf{N}_{lm}(k,\mathbf{r}-\mathbf{r}_c)\cdot b_{lm} + \mathbf{L}_{lm}(\kappa,\mathbf{r}-\mathbf{r}_c)\cdot c_{lm},$$
$$\mathbf{H}(\mathbf{r}-\mathbf{r}_c) = \frac{1}{iZ}\sum_{lm}\mathbf{N}_{lm}(k,\mathbf{r}-\mathbf{r}_c)\cdot a_{lm} + \mathbf{M}_{lm}(k,\mathbf{r}-\mathbf{r}_c)\cdot b_{lm}. \tag{12}$$

In Eq. (12), $k$ is the transverse wavenumber and $\kappa$ is the longitudinal wavenumber; and $\mathbf{r}$ is a field point in space, $\mathbf{r}_c$ is the center of an interface. The summation in (12) is conducted with respect to $l$ and $m$, where $l$ is a nonnegative integer and $m$ is an integer between $-l$ and $+l$. $Z$ is the transverse wave impedance. $a_{lm}$, $b_{lm}$ and $c_{lm}$ are the expansion coefficients of the spherical wave functions whose expressions are [72],

$$\mathbf{M}_{lm}(k,\mathbf{r}-\mathbf{r}_c) = z_l(kr)\cdot\mathbf{X}_{lm}(\theta,\varphi),$$
$$\mathbf{N}_{lm}(k,\mathbf{r}-\mathbf{r}_c) = l(l+1)\frac{z_l(kr)}{kr}\cdot Y_{lm}(\theta,\varphi)\hat{\mathbf{r}} + \frac{1}{kr}\frac{\partial(rz_l(kr))}{\partial r}\cdot\mathbf{Z}_{lm}(\theta,\varphi),$$
$$\mathbf{L}_{lm}(k,\mathbf{r}-\mathbf{r}_c) = \frac{\partial z_l(\kappa r)}{\partial(\kappa r)}\cdot Y_{lm}(\theta,\varphi)\hat{\mathbf{r}} + \frac{z_l(\kappa r)}{\kappa r}\cdot\mathbf{Z}_{lm}(\theta,\varphi). \tag{13}$$

Here, the subscripts $l$ and $m$ are referred to as the degree and order of the spherical harmonics $Y_{lm}(\theta,\varphi)$ (see Appendix D in [72]). $z_l$ is the spherical Bessel function or the spherical Hankel function, depending on whether the wave has a standing wave or radiating wave nature. The vector spherical harmonics, i.e., $\mathbf{X}_{lm}(\theta,\varphi)$ and $\mathbf{Z}_{lm}(\theta,\varphi)$, are,

$$\mathbf{X}_{lm}(\theta,\varphi)=\frac{1}{\sin\theta}\frac{\partial Y_{lm}(\theta,\varphi)}{\partial\varphi}\hat{\boldsymbol{\theta}}-\frac{\partial Y_{lm}(\theta,\varphi)}{\partial\theta}\hat{\boldsymbol{\varphi}},$$
$$\mathbf{Z}_{lm}(\theta,\varphi)=\frac{\partial Y_{lm}(\theta,\varphi)}{\partial\theta}\hat{\boldsymbol{\theta}}+\frac{1}{\sin\theta}\frac{\partial Y_{lm}(\theta,\varphi)}{\partial\varphi}\hat{\boldsymbol{\varphi}}. \tag{14}$$

In the above, $\hat{\mathbf{r}}$, $\hat{\boldsymbol{\theta}}$ and $\hat{\boldsymbol{\varphi}}$ are spherical unit vectors of a spherical coordinate system centered at $\mathbf{r}_c$, shown in Fig. 2.

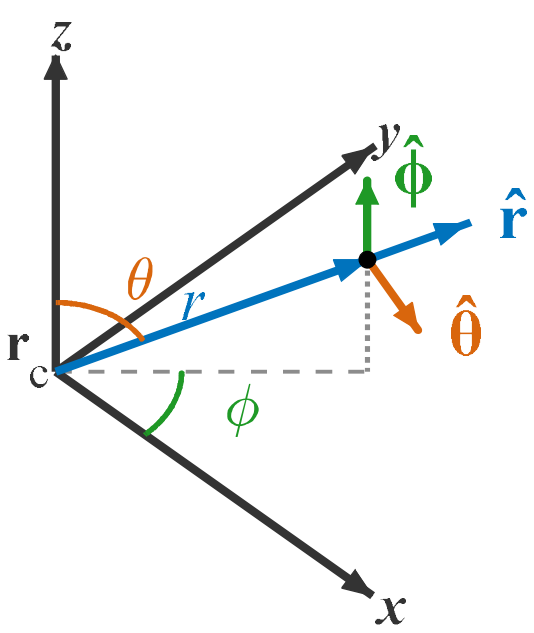

Fig. 2. Schematic of the spherical coordinate system. The origin $\mathbf{r}_c$ is located at the center of the NP. A spatial point is defined by $(r,\theta,\varphi)$ with the corresponding local orthonormal unit vectors $\hat{\mathbf{r}}$, $\hat{\boldsymbol{\theta}}$ and $\hat{\boldsymbol{\varphi}}$.

Then, we focus on the construction of the **S** matrix. The **S** matrix connects the expansion coefficients of all outgoing spherical waves with all incoming spherical waves that leave or enter, respectively, an interface. Its construction has been introduced in our previous work [43,66]. First, as shown in Eq. (12) and (13), the electric and magnetic fields are expanded in terms of spherical wave functions. Second, the expanded fields are matched through boundary conditions. By projecting both sides of the boundary conditions discussed in Section 2.2 onto $\mathbf{X}_{lm}(\theta,\varphi)$ and $\mathbf{Z}_{lm}(\theta,\varphi)$ (and $Y_{lm}(\theta,\varphi)\hat{\mathbf{r}}$, if HDM is considered), the projection matrices (referred to as the **M** matrices) are constructed. Then, the algorithm behind the toolbox is divided into two systems, distinguishing the transverse electric (TE) wave system from the transverse magnetic (TM) wave system. Depending on the mesoscopic physical model used, the number of boundary conditions varies. For a local-local interface, under the LRA two boundary conditions in Eq. (6) are considered; while for SRM, the quantum-corrected boundary conditions Eq. (10) and Eq. (11) are applied instead. For a local-nonlocal interface or a nonlocal-local interface, three boundary conditions in Eq. (6) and Eq. (8) must be satisfied. For a nonlocal-nonlocal interface, in total four equations in Eq. (6) and Eq. (9) are required. The explicit and detailed expressions for the **M** matrices can be found in [43]. Such a projection leads to

$$\mathbf{M}_1^+\cdot w_1^+ +\mathbf{M}_1^-\cdot w_1^- =\mathbf{M}_2^+\cdot w_2^+ +\mathbf{M}_2^-\cdot w_2^-. \tag{15}$$

Here, $w$ is a collection of expansion coefficients and is defined as a column vector $w = [a_{lm}; b_{lm}; c_{lm}]$ where $c_{lm}$ is added if HDM is used, and each of $a_{lm}$, $b_{lm}$, $c_{lm}$ is a representative element of a column vector with the size of $(l_{\max}+1)^2$. The superscripts – or + indicate whether the waves are propagating against or along the radial direction, and the subscripts 1 or 2 indicate the inner or the outer region. Eq. (15) can be written as,

$$\begin{bmatrix} w_1^- \\ w_2^+ \end{bmatrix}=\begin{bmatrix}\mathbf{M}_1^- & -\mathbf{M}_2^+\end{bmatrix}^{-1}\cdot\begin{bmatrix}-\mathbf{M}_1^+ & \mathbf{M}_2^-\end{bmatrix}\begin{bmatrix} w_1^+ \\ w_2^- \end{bmatrix}. \tag{16}$$

Then, an **S** matrix can be obtained by

$$\mathbf{S}=\begin{bmatrix}\mathbf{M}_1^- & -\mathbf{M}_2^+\end{bmatrix}^{-1}\cdot\begin{bmatrix}-\mathbf{M}_1^+ & \mathbf{M}_2^-\end{bmatrix}. \tag{17}$$

## 2.4 Translation Matrix

As noted in Eq. (12), the expansion of the electric and the magnetic field is with respect to the center of each interface. Given that the fields are unique, it is important to link the spherical wave functions expanded regarding different interface centers. This spatial translation is elegantly achieved via the translational addition theorem [72]. Spherical wave functions expressed relative to $r_c$ (the expansion center of the observation interface $p$) and $r_c^{'}$ (the expansion center of the source interface $q$) can be linked through

$$\begin{aligned}
\mathbf{M}_{l'm'}(\mathbf{r}-\mathbf{r}_c^{'}) &= \sum_{lm}[\mathbf{M}_{lm}(\mathbf{r}-\mathbf{r}_c)\cdot A_{lm,l'm'}(\mathbf{r}_c^{'}-\mathbf{r}_c)+\mathbf{N}_{lm}(\mathbf{r}-\mathbf{r}_c)\cdot B_{lm,l'm'}(\mathbf{r}_c^{'}-\mathbf{r}_c)],\\
\mathbf{N}_{l'm'}(\mathbf{r}-\mathbf{r}_c^{'}) &= \sum_{lm}[\mathbf{M}_{lm}(\mathbf{r}-\mathbf{r}_c)\cdot B_{lm,l'm'}(\mathbf{r}_c^{'}-\mathbf{r}_c)+\mathbf{N}_{lm}(\mathbf{r}-\mathbf{r}_c)\cdot A_{lm,l'm'}(\mathbf{r}_c^{'}-\mathbf{r}_c)],\\
\mathbf{L}_{l'm'}(\mathbf{r}-\mathbf{r}_c^{'}) &= \sum_{lm}\mathbf{L}_{lm}(\mathbf{r}-\mathbf{r}_c)\cdot C_{lm,l'm'}(\mathbf{r}_c^{'}-\mathbf{r}_c).
\end{aligned} \tag{18}$$

The translation matrix is a block matrix composed of $A$ and $B$ matrices (and $C$ matrices, if the nonlocal response model is used). The explicit analytical expression of the elements in the $A, B$ and $C$ matrices can be found in Appendix D of [72], and the translation matrix **T** is compactly represented by

$$\mathbf{T}_{p,q}=\begin{bmatrix}A_{lm,l'm'}\left(k,\mathbf{r}_c^{'}-\mathbf{r}_c\right) & B_{lm,l'm'}\left(k,\mathbf{r}_c^{'}-\mathbf{r}_c\right) & 0\\ B_{lm,l'm'}\left(k,\mathbf{r}_c^{'}-\mathbf{r}_c\right) & A_{lm,l'm'}\left(k,\mathbf{r}_c^{'}-\mathbf{r}_c\right) & 0\\ 0 & 0 & C_{lm,l'm'}\left(\kappa,\mathbf{r}_c^{'}-\mathbf{r}_c\right)\end{bmatrix}. \tag{19}$$

In Eq. (19), $A_{l'm',lm}$ and $B_{l'm',lm}$ are the representative elements of matrix blocks that describe the translation of the transverse vector wave functions from one expansion center to another (thus they are dependent on the transverse wave number $k$), while $C_{l'm',lm}$ being for the translation of the longitudinal vector wave functions (thus it is a function of the longitudinal wave number $\kappa$).

Physically, the scattered field originating from the $q$th interface, characterized by its expansion coefficients $a_{l'm'}$ , $b_{l'm'}$ and $c_{l'm'}$ , acts as a secondary incident field impinging on the $p$th interface. Then, the secondary incident field must be expanded in terms of vector spherical wave functions centered at $r_c$. Together with Eq. (18), the expansion coefficients with respect to the center of the $q$th interface can be linked with the coefficients with respect to the center of the $pth$ interface via the **T** matrix in Eq. (19),

$$\begin{bmatrix}a_{lm}\\ b_{lm}\\ c_{lm}\end{bmatrix}=\mathbf{T}_{p,q}\cdot\begin{bmatrix}a_{l'm'}\\ b_{l'm'}\\ c_{l'm'}\end{bmatrix}. \tag{20}$$

## 2.5 Main Equation

First, we focus on the $w$th interface (here we assume that there are $N$ interfaces in total and thus $w$ is an integer spanning from 1 to $N$). For the $w$th interface, the **S** matrix that links the expansion coefficients of all the incoming/outgoing waves is

$$\begin{pmatrix} w_1^- \\ w_2^+ \end{pmatrix} = \mathbf{S}_w \cdot \begin{pmatrix} w_1^+ \\ w_2^- \end{pmatrix}. \tag{21}$$

The right-hand side (RHS) of Eq. (21) represents the expansion coefficients of all the incoming fields. In particular, the RHS consists of two parts. The first part regards the direct incident field, marked as $w^{\text{exc}}$; the second part comes from the scattered fields of all the other interfaces $q$, linked by the translation matrix. Therefore, the total incident field can be expressed as,

$$\begin{pmatrix} w_1^+ \\ w_2^- \end{pmatrix} = w^{exc} + \sum_q \begin{pmatrix} \mathbf{T}_{w,q}(+,1,-,1) & \mathbf{T}_{w,q}(+,1,+,2) \\ \mathbf{T}_{w,q}(-,2,-,1) & \mathbf{T}_{w,q}(-,2,+,2) \end{pmatrix} \cdot \begin{pmatrix} q_1^- \\ q_2^+ \end{pmatrix}. \tag{22}$$

The subscript "$w, q$" says that the matrix $\mathbf{T}$ describes how the scattered field emitted by the $q$th interface (as a "source" interface) is seen by the $w$th interface (as the "observation" interface). The 1st and 2nd arguments of $\mathbf{T}_{w,q}$ in the brackets originate from the super- and subscripts of the expansion coefficients at the $w$th interface, while the 3rd and 4th arguments in the brackets are from the super- and subscripts of the expansion coefficients at the $q$th interface. Then, we put Eq. (22) into Eq. (21) and shuffle the terms on both sides, with the main equation for the $w$th interface finally being listed as

$$\begin{pmatrix} w_1^- \\ w_2^+ \end{pmatrix} - \mathbf{S}_w \cdot \sum_q \begin{pmatrix} \mathbf{T}_{w,q}(+,1,-,1) & \mathbf{T}_{w,q}(+,1,+,2) \\ \mathbf{T}_{w,q}(-,2,-,1) & \mathbf{T}_{w,q}(-,2,+,2) \end{pmatrix} \cdot \begin{pmatrix} q_1^- \\ q_2^+ \end{pmatrix} = \mathbf{S}_w \cdot w^{exc}. \tag{23}$$

In this main equation, the expansion coefficients of the scattered fields are what we want to solve. The same equation can be formulated for all the other $N - 1$ interfaces. In this way, we obtain $N$ equations in total with $N$ unknowns. By solving these equations, together with the excitation field, the electric field can be reconstructed in the whole space.

# 3 The Toolbox

## 3.1. Conventions

For clarity and consistency, the proposed toolbox adheres to the following physical conventions and numerical specifications:

1.Supported Excitations: the excitation can be chosen among arbitrarily polarized plane waves, dipoles, and electron beams.

2.Background Medium: the background of the plasmonic scatterer is always assumed to be a homogeneous and lossless medium.

3.Unit System: The International System of Units is adopted throughout the software.

4. Spatial and Angular Dimensions: All the geometric parameters and wavelengths are defined in nm. While the solver processes angles in radians (rad), the inputs in all demos (see below) are entered in degrees (°) and translated to rad.

This toolbox is implemented in MATLAB and users are assumed to be familiar with the environment and syntax of it.

## 3.2. Installation

The proposed toolbox is distributed as three compressed archives together with this paper. Users can download and extract all three files into a local directory of their choice, then add this folder along with its subfolders to the MATLAB search path. Once the path is set, the given demo scripts can be executed directly within MATLAB. The proposed toolbox is self-sufficient and implemented for the standard MATLAB environment, operating seamlessly without relying on any additional proprietary MATLAB toolboxes. Nonetheless, we used meshes generated from an open source mesher *Gmsh*, for the mapping of far- and near-field distributions, as well as the calculation of optical cross sections, radiated and absorbed power, decay rates, EEL and CL probabilities. Of course, any implementation that follows the data format functions perfectly.

### 3.3. Structure

The proposed toolbox consists of various directories and subdirectories to facilitate the reading of routines of different functionalities and level of organization. The result of this choice is a rather involved file system. Instead of providing a tree diagram, that would be rather cumbersome and uninformative, we present the main directories in Table 1. After browsing through the folders, the names and functionalities of each routine are elucidated in Section 3, and in the comments of each routine.

Table 1. First level organization of the proposed toolbox.

| Directory | Functionality |
| --- | --- |
| Main/0_T | Contains the functions to calculate the S matrix of each interface, and the translational matrices relating interfaces centered at different positions. |
| Main/1_exc | Contains the functions setting up excitations, including 1) plane wave excitation, 2) dipole excitation and 3) electron beam excitation. |
| Main/2_solve | Contains the functions that solve the linear system as in Eq. (23). |
| Main/3_flds | Contains the postprocessing functions, e.g., the ones that calculate the electric and magnetic fields in space, the absorption and scattering cross sections, near-field, far-field mapping, charge and chirality. |
| Misc/0_material | Contains material models and commonly used material data from [73], 1) for LRA; 2) for HDM, 3) for SRM. |
| Misc/1_mathfun | Contains basic math functions, including factorial, square root, multidimensional matrix multiplication [74], supplemented by open-source utilities such as facted, gamma0, fact (by Paul Godfrey), multinv (by Xiaodong Qi) multiprod and multitransp (by Paolo de Leva) under their respective open-source licenses; vector spherical wave functions; spherical Bessel and Hankel functions; and associated Legendre polynomials. |
| Misc/2_mesh | Contains the example meshes generated by the software *Gmsh*, for post-processing. |
| Misc/3_particles | Contains all the functions aiming to generate an entity which contains all the necessary information for each interface. In detail, the |

| | information includes the geometrical and material characteristics, as well as the S matrix of each interface. |
|---|---|
| Misc/4_waitbar | Contains a simple wait bar to visualize the estimated required runtime for the simulations, supplemented by open-source utilities multiWaitbar (by Ben Tordoff) under its respective open-source licenses. |

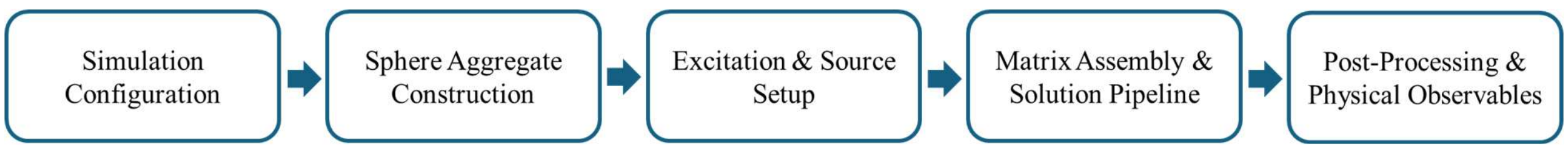

Fig. 3. The workflow of the proposed MATLAB toolbox.

To provide a global overview of the proposed computational framework, the workflow of the proposed MATLAB toolbox is delineated in Fig. 3. The architecture is modularly partitioned into five sequential pipelines:

(i) **Simulation Configuration**, where global simulation parameters, numerical tolerances, and the look-up table for the translation matrix are initialized via `simulation.m`;

(ii) **Structure Modeling**, where the spatial geometry and the material of the sphere aggregates are defined using `particle.m`, and link between different spherical interfaces is defined using `multiparticletab.m`;

(iii) **Source Modeling**, where the incident field characteristics are established via `multiexc.m`;

(iv) **Toolbox Execution**, wherein `multisolve.m` handles the $\mathbf{S}$-matrix assembly, the evaluation of the translational addition theorems and the solution of the main equation; and

(v) **Post-Processing**, where various routines are provided: scattered EM field and scattering cross sections are calculated via `fldmapping.m`; absorption cross sections are calculated via `absorption.m`; `fluoenhancement.m` and `quantum.m` are invoked to extract molecule excitation rate and quantum yield, and the EEL probability and CL probability are evaluated in `eels.m` and `cl.m`, respectively.

In the following, we elaborate on each step.

### 3.4. Inputs

The toolbox operates via main scripts tailored to individual application scenarios. Users do not need to construct these scripts from scratch, as new scripts can be created by modifying the existing templates provided. To illustrate the required input parameters, this section dissects and explains an input demo as the first part of the main script. Since the excitation has three different ways: plane waves, dipoles and electron beams, the three excitation setups are given in detail.

#### *3.4.1 Basic Parameters*

Before the setup of the basic parameters, the translation coefficient tables used to construct the **T** matrix must be initialized. Users only need to run the `init_coeffs.m` file to complete this step. This initialization program pre-generates all spherical harmonic translation coefficients required by the wave expansion and stores them in a block-structured database. All example demos already include these translation coefficient tables and an additional summary file `Tcoeffs.mat` records the $l$-range of each group. In the current setup, the upper limit for `lmax` is 29, which is sufficient for virtually all application scenarios. If a larger upper limit is required, the corresponding parameters in `init_coeffs.m` should be adjusted, where the needed translation coefficient tables will be regenerated automatically (please refer to Appendix A for more information).

First, the wavelengths are included in a row vector `enei`,

```
% enei ( wavelength )

enei = linspace( 300, 700, 41 );
```

Notice that `enei` spans 41 values between 300 and 700 nm, the default units. The users then define the highest $l$ order, i.e., `lmax`, at which the expansions in spherical wave functions are truncated. Below, `lmax` is set to 19. `lmax` is the critical convergence parameter and this choice cannot be arbitrary. Users can perform a convergence test before trusting their data, picking `lmax` as the lowest order which satisfies a tolerance criterion with respect to analytical, precise numerical or next step simulations.

```
% lmax

lmax = 19;
```

Other basic parameters for the simulation are defined in `simulation.m`, including physical constants (permittivity, permeability, vacuum impedance, Planck's constant and electron Volt), default units (wavelength and frequency unit), the orders of spherical wave functions, frequency parameters (angular frequency and wavenumber), the look-up table for the calculation of translation matrix, solution parameters, and simulation tolerances. These are neatly returned in a simulation struct, which will be a standard input to all our routines below.

### *3.4.2 Spherical Interfaces*

First, the materials to be used in the toolbox are introduced, both for the scatterers and the surrounding environment. This includes three parts. To start with, the EM response of bulk materials used in the simulation is defined through a *cell* variable `sim.bulk`,

```
% bulk

sim.bulk = { materialconst(          3, sim );...

             materialnlocal( 'Au', 'LD', sim );...

             materialconst(          1, sim );...

              };
```

Here, the `sim.bulk` variable includes three elements corresponding to a material with a *constant* relative permittivity $\varepsilon_r = 3$ (this is why the `materialconst` function is used and, by *constant*, we mean the relative permittivity is nondispersive); a *nonlocal* material whose EM response is described by the HDM (this is why the `materialnlocal` function is used); and again a material with a *constant* relative permittivity $\varepsilon_r = 1$. In the toolbox, the three media are named as *medium 1*, *medium 2* and *medium 3*. As a remark, for *medium 2*, `'Au'` may be replaced by Ag, Al or Na so that the EM response of silver, aluminum or sodium can be modelled. Also, `'LD'` refers to the Lorentz – Drude model [73] while `'D'` can be used and refers to the Drude model. Lastly, similar to the `materialnlocal` function, the `materiallocal` function can be found in the toolbox and gives the conventional LRA model.

Further, since the SRM is included in the model, the possibility of setting up the quantum-corrected BC is done through,

```
% surf
sim.surf = { dpara( 'Null', sim ) };
```

The `sim.surf` variable is a *cell* variable, which includes the surface models used in the simulation. Each entry of the cell variable is numbered and the number is later referred to identify which surface model is used (see more details along our discussions on the setup of `particle.m`. In the above, since in the simulation of concern the SRM is not considered, the first argument of the `dpara` function is set to `'Null'`. However, in case SRM is used, since the proposed solver only supports Na currently, the first argument of the `dpara` function is set to `'Na'`. The $d$ parameters of other noble metals can be implemented through the analytical approach in [75].

```
% surf
sim.surf = { dpara( 'Na', sim ) };
```

Last, since the toolbox analyzes the scattering of multiple spherical objects in a homogeneous background, the background medium needs to be specified. It is set through the following line.

```
% env ( in which medium the spheres are )
sim.env = 3;
```

This line says that *medium 3*, that is, the medium with a constant relative permittivity $\varepsilon_r = 1$, is the background medium (the flag given above corresponds to the last element of `sim.bulk`; this is a deliberate choice of the user and not a convention of the solver).

As indicated in Section 2.5, the toolbox is based on the concept of interfaces that separate one bulk medium from another. Here, we discuss three potential configurations (see Fig. 4). The first case is a concentric core-shell structure (see Fig. 4(a)).

```
% particle 1 ( center, radius, in/out materials, on-boundary material )
rc1 = [ 0, 0, 0 ]; r1 = 25; inout1 = [ 1, 2 ]; on1 = 1;
```

In the above, for the first line, the geometric parameters are defined for the first interface. `rc1` indicates the origin, `r1` defines the radius of 25 nm. `inout1` indicates that for interface 1, the inner region is filled with *medium 1* (the first parameter) and outer region is filled with *medium 2* (the second parameter); the flag `inout` reflects the position the user has defined in the `sim.bulk` (material) cell array. `on1` decides the on-boundary material (whether to use the SRM), indicating the first cell in `sim.surf`; in general the flag `on` reflects the position the user has defined in the `sim.surf` (Feibelman) cell array. A similar set up can be done for interface 2,

```
% particle 2 ( center, radius, in/out materials, on-boundary material )
rc2 = [ 0, 0, 0 ]; r2 = 85; inout2 = [ 2, 3 ]; on2 = 1;
```

The above configuration is shown in Fig. 4 (a). The pink part represents *medium 1*, and the light blue region is filled with *medium 2*, while the background is filled with *medium 3*.

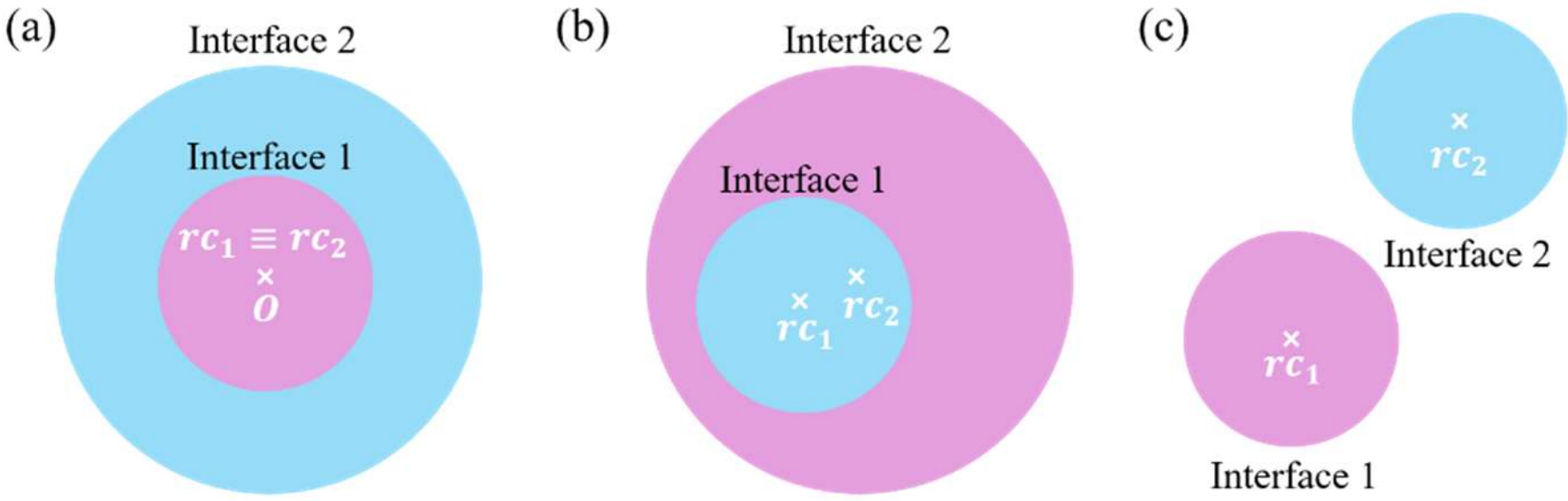

Fig. 4. Illustration of three possible two-interface configurations. (a) Interface 1 is inside of Interface 2, with their centers located at the origin. (b) Interface 1 is inside Interface 2, with different centers. (c) Interface 1 and 2 are adjacent but not in contact, owning different centers.

The second configuration is illustrated in Fig. 4(b). It is an eccentric nanosphere. Thus, the geometry and the material configuration are defined through,

```
% particle 1 ( center, radius, in/out materials, on-boundary material )
rc1 = [ -5, -3, 0 ]; r1 = 25; inout1 = [ 2, 1 ]; on1 = 1;
```

Here, `rc1` indicates the spatial position of the origin, `r1` defines the radius which is 25 nm. `inout1` indicates that for interface 1, the inner region is filled with *medium 2* and outer region is filled with *medium 1*. `on1` decides the on-boundary material (whether to use the SRM). Here, since `on1` is set to 1, the first cell in `sim.surf` is referred to. A similar set up can be done for interface 2,

```
% particle 2 ( center, radius, in/out materials, on-boundary material )
rc2 = [  1,  0, 0 ]; r2 = 85; inout2 = [ 1, 3 ]; on2 = 1;
```

The third configuration is shown in Fig. 4(c). The two nanospheres are put adjacent.

```
% particle 1 ( center, radius, in/out materials, on-boundary material )
rc1 = [ -30, -40, 0 ]; r1 = 25; inout1 = [ 1, 3 ]; on1 = 1;
```

This line sets the geometric parameters for the first nanosphere. It is origin at (-30 nm, -40 nm, 0 nm), with a radius of 25 nm. `inout1` indicates that for interface 1, the inner region is filled with *medium 1* and outer region is filled with *medium 3*. `on1` decides the on-boundary material, indicating the first cell in `sim.surf`. A similar set up can be done for interface 2,

```
% particle 2 ( center, radius, in/out materials, on-boundary material )
rc2 = [ +30,  40, 0 ]; r2 = 25; inout2 = [ 2, 3 ]; on2 = 1;
```

Then, these parameters are fed into the `particle` function which sets up the **S** matrix (see Eq. (17)) for each particle,

```
% particle 1
```

```
p1 = particle( rc1, r1, inout1, on1, sim );
% particle 2
p2 = particle( rc2, r2, inout2, on2, sim );
```

After setting up each interface, the interaction between interfaces is accounted for via the `multiparticletab` function,

```
% p
p = multiparticletab( { p1, p2 }, sim );
```

If there is a displacement between two spherical interfaces (that is, $\mathbf{r}_{c1} \neq \mathbf{r}_{c2}$), the $\mathbf{T}$ matrix (see Eq. (19)) is evaluated by this function.

*3.4.3 Excitations*

*A. Plane Waves*

The plane wave excitation is defined by three variables, i.e., `theta`, `phi` and `pol`,

```
% plane wave
theta = linspace( 0, 90, 10);
phi = linspace( 0, 0, 1);
pol = 'TM';
```

`theta` and `phi` refer to the $\theta$ and the $\varphi$ angles as in a spherical coordinate system (Fig. 2, `rc` is origin at (0 nm, 0 nm, 0 nm) in this case). The propagation direction of the plane wave is along the radial direction of the system (see the unit vector $\hat{\mathbf{r}}$ in Fig. 2), with $\hat{\boldsymbol{\theta}}$ being the direction of the electric field of a TM wave (set up by `pol = 'TM'`) and $\hat{\boldsymbol{\varphi}}$ being the direction of the electric field of a TE wave (set up by `pol = 'TE'`). The reference point of the plane wave is set at the origin of coordinate system (see Fig. 2). In the above, $\theta$ spans from 0 to 90 degrees with 10 degrees as a step and $\varphi$ is 0 degree. To be complete, besides the TM and the TE waves, users can choose left circular polarization (LCP) and right circular polarization (RCP) as well. The definition of the LCP and the RCP waves follows the convention in [76]. Last, the three variables are fed into the `multiexc` function to set up the excitation of the system. The amplitude of the plane wave is assumed unit (though this can be edited as well).

```
% p excitation
p = multiexc( p, sim, theta, phi, pol, 'pw' );
```

*B. Dipoles*

For the dipole excitation, the positions and moments of the dipoles are needed.

```
% pos ( positions of dipoles )
pos = [ 90, 0, 0; ...
        0, 90, 0; ...
        0, 0, 90; ...
```

```
        0, 0, 90 ];
% pol ( dipole moment )
pol = [ 0, 0, 1;
       0, 0, 1;
       0, 0, 1;
        1/sqrt( 2 ), 0, 1/sqrt( 2 )];
```

The information is stored in `pos` and `pol` variables. The two variables are of the same size. The size is the number of dipoles multiplied by 3. For `pos (pol)`, the second dimension marks the $x$, $y$, $z$ coordinates (components) of the displacement vector (dipole moments). In the example above, we consider 4 dipoles whose positions are at (90 nm, 0 nm, 0 nm), (0 nm, 90 nm, 0 nm), (0 nm, 0 nm, 90 nm) and (0 nm, 0 nm, 90 nm), respectively, and the corresponding dipole moments are along $\hat{z}$, $\hat{z}$, $\hat{z}$ and $1/\sqrt{2}\,\hat{x}+1/\sqrt{2}\,\hat{z}$. The `pos` and `pol` variables are fed into the `multiexc` function to set up the excitation of the system,

```
% p excitation
p = multiexc( p, sim, pos, pol, 'dip' );
```

*C. Electron Beams*

For the electron beam excitation, we require the configuration of the positions and the speed of electrons.

```
% r
r = [100, 0, 0; ...
   125, 0, 0 ];
% speed
v = [0.33; 0.33] * sim.c;
```

This information shows the excitation position and the speed of electrons. The size of `r` variable is the number of electron beams multiplied by 3, and the second dimension displays the $x$, $y$, and $z$ coordinate components. In this demo, the two electron beams are emitted from (100 nm, 0 nm, 0 nm) and (125 nm, 0 nm, 0 nm) at one-third the speed of light, respectively. `r` and `v` are fed into the `multiexc` function to set up the excitation of the system,

```
% p excitation
p = multiexc( p, sim, r, v, 'eels' );
```

### 3.5. Solver

After setting up the spherical interfaces and excitations, we are ready to solve the linear system in Eq. (23),

```
% p solve
p = multisolve( p, sim );
```

As we have mentioned before, the user may opt between direct and iterative solvers by editing the `sim.sol` field in the `simulation` routine. Due to significant scaling the **S** and **T** matrices manifest, on account of the coexistence of transverse and longitudinal contributions [43], we equilibrate the system matrix before inversion.

### 3.6. Post Processing

After solving the main equation, we are ready to evaluate the scattered field due to multiple spherical interfaces.

*3.6.1 Near Field and Far Field Calculation*

On the one hand, the calculation of near fields can be done as,

```
% nf

nf = nffldmapping( 'demo5050nf', p, sim );
```

The sample mesh file `demo5050nf.m` is a 50 nm * 50 nm square in the $xy$ plane, centered at (0 nm, 10 nm, 0 nm), generated by *Gmsh*, with 23382 discretized triangles in total and the maximum radius of the inscribed circle of the triangular elements is 0.2365 nm. `nf` is a struct, where `nf.E` and `nf.H` store the total electric and magnetic fields at the incenters of the triangles in the mesh.

On the other hand, the calculation of the far fields is done as through,

```
% load mesh

ff = ffsphere;

ff = fldpts( ff, p, sim);

% the electric field and magnetic field

[ ff.E, ff.H ] = fldmapping( ff, p, sim );
```

The meshes `ff` is generated by *Gmsh*, with 4232 discretized triangles in total and the maximum inscribed circle radius is 0.0369 m. `ff.E` and `ff.H` store the scattered electric and magnetic fields at the incenters of the far field mesh.

Both far and near fields are simply reconstructed through the vector spherical wave expansions in Eq. (12) and given the calculated coefficients.

After clarifying the electric and magnetic field in the near and the far field regime, it is time to carry out the post processing steps to get the desired physical parameters. In the following, we discuss: (1) the calculation of optical cross sections; (2) the calculation of Electron Energy-Loss (EEL) and Cathodoluminescence (CL) probabilities; (3) the calculation of fluorescence enhancement, quantum yield and Purcell factor. Three corresponding examples are given in Section 4.

*3.6.2 Optical Cross Sections*

First, the scattering/absorption cross sections are calculated by integrating the Poynting vector (of the scattered/total field) over the sphere defined by `ffsphere` or a sphere just outside each spherical object (which may include several concentric or non-concentric spherical interfaces). The calculations of the scattering and absorption cross sections are in the following lines:

```
% scattering
p = ffmapping( p, sim );
% absorption
p = absorption( p, sim );
```

The scattering cross sections and the corresponding scattering power can be found in `p.ff.sca` and `p.ff.psca`; while the absorption cross sections and absorbed power are in `p.nf.abt` and `p.nf.pabt`.

*3.6.3 Cathodoluminescence (CL) and Electron Energy Loss (EEL)*

For electron beam excitations, the EEL and CL can be calculated. We assume that 1) the electrons cannot penetrate through the sample and 2) the electrons travel along the *z* axis. On the one hand, CL is closely related to the radiated power. Thus, once again it can be evaluated as the surface integral of the Poynting vector of the scattered field on the sphere defined by `ffsphere` by the `cl` function (of course, within a normalization constant [54]),

```
% CL S Matrix Method
CL0 = cl( ff, p, sim);
```

On the other hand, EEL is the retarded work performed by the induced electric field along the electron's trajectory (see the detailed formula leading to EEL in [54]) and is calculated by the `eels` function,

```
%% EELS
% scattering
p = eels(p, r, sim);
% EELS S Matrix Method
EELS0 = p.eels ./ sim.ev;
```

*3.6.4 Fluorescence Enhancement, Quantum Yield and Purcell Factor*

The proposed toolbox is also able to calculate the fluorescence enhancement and emitter quantum yield, treating a fluorescent molecule as an electric dipole. This is known to be an accurate representation of the transitions of a realistic molecule down to 1-nm gaps, before the spatial extent of the molecule causes deviations [77]. In the first stage, the metallic NP is excited by plane waves, the resulted enhanced electric field excites the molecule, resulting in increased excitation rate. As the normalized excitation rate $\gamma_{exc} \propto |p_d \mathbf{E}(R_d)|^2$. $p_d$ is the dipole moment and $R_d$ is the *r* component of the dipole position. The codes given below show the calculation, and the information of particles `p` and the simulation parameters `sim` are fed into the `fluoenhancement` function.

```
% the normalized excitation rate
eta = fluoenhancement( p, sim);
```

Here, the normalized excitation rate is stored at `eta`.

In the second stage, the molecule radiates, where the Purcell effect induced by the metallic NP alters both the radiative and non-radiative power of the molecule, thereby modifying its quantum yield [78]. While the Purcell factor is calculated as the scattered power `p.purcell.sumpsca` over a closed surface containing the nanostructure and the dipole (with the radius equal to 1.001× the radius of the sphere containing the dipole ), divided by the corresponding power of the dipole alone `p0`, the quantum yield is obtained as the division of the radiated power `p.ff.psca` and sum of radiated power `p.ff.psca` and non-radiative power `p.nf.pabt` [79].

$$\frac{q}{q^0} = \frac{P_{rad}}{P_{rad} + P_{abt}} \tag{24}$$

```
% quantum yield & Purcell Factor

[quany, purcellS] = quantum(p, sim);
```

`quany` represents the calculated normalized quantum yield and `purcellS` represents the obtained Purcell factors. Afterwards, the fluorescence enhancement can be calculated as the product of normalized excitation rate and quantum yield, using the equation implied in [80].

$$\frac{\gamma_{em}}{\gamma_{em}^0} = \frac{\gamma_{exc}}{\gamma_{exc}^0} \cdot \frac{q}{q^0} \tag{25}$$

```
% fluorescence enhancement

yemS = quany .* eta;
```

Here, the fluorescence enhancement is stored at `yemS`.

## 4. Examples

### 4.1 Fluorescence Enhancement

In the first example, we consider the fluorescence enhancement of a dipole emitter near a concentric NP. The simulation is a two-step one. For the first step, we choose a TM polarized plane wave as excitation. A dipole, as a fluorescent molecule, is placed on top of the NP, whose dipole orientation is fully aligned with the polarization direction of the incident electric field. Such a configuration promises the generation of the highest fluorescence rate [79,81]. For the second step, the dipole acts as a secondary source, and the radiative power and non-radiative power of the molecule are modified by the presence of the NP. The full simulation set-up can be found in `example1.m` provided in the source package.

For this example, the users define the simulated frequency band, the geometry, as well as the detailed excitation. First, we choose a wavelength range from 300 nm to 900 nm with 61 points taken in between and the maximum harmonic order `lmax` is set to 12 (as a relatively large concentric nanosphere is considered).

```
% enei ( wavelength )

enei = linspace( 300, 900, 61 );

% lmax

lmax = 12;
```

We consider a concentric nanosphere, shown in Fig. 5(a), whose core is made from silver (*medium 1*), and the shell is made from gold (*medium 2*) [82]. The metallic material is treated as a nonlocal medium using HDM and the bulk properties are described by the Lorentz-Drude Model, indicated by `'Ag'`, `'Au'` and `'LD'` set in the `materialnlocal` function. Since SRM is not considered here, the first parameter in the cell `sim.surf` is set as `'NULL'`. The background of nanostructure is indicated by `sim.env`, while here 3 represents *medium 3* (the dielectric material with relative permittivity equal to 1, vacuum).

```
%% materials

% bulk

sim.bulk = {

             materialnlocal( 'Ag', 'LD', sim );...

             materialnlocal( 'Au', 'LD', sim );...

             materialconst(          1, sim );...

              };

% surf

sim.surf = { dpara( 'Null', sim ) };

% env ( in which medium the spheres are )

sim.env = 3;
```

Then we set the geometric parameters. The radius of the silver core `r1` is 25 nm, and the radius of the whole nanosphere `r2` is 85 nm. These two nanospheres are centered at the origin. The `inout1` parameter regulates the inner region filled with *medium 1* and outer region filled with *medium 2*, while `inout2` indicates that the inner region is filled with *medium 2* and outer region is filled with *medium 3.* Since SRM is not considered in this demo, `on1` and `on2` are both set to 1, indicating the first cluster defined in `sim.surf`.

```
%% particles

% particle 1 ( center, radius, in/out materials, on-boundary material )

rc1 = [ 0, 0, 0 ]; r1 = 25; inout1 = [ 1, 2 ]; on1 = 1;

rc2 = [ 0, 0, 0 ]; r2 = 85; inout2 = [ 2, 3 ]; on2 = 1;
```

After that, the `particle` function follows.

```
% p1 & p2

p1 = particle( rc1, r1, inout1, on1, sim );

p2 = particle( rc2, r2, inout2, on2, sim );
```

Then, for the plane wave excitation as the first stage, we used a TM plane wave propagating along the $x$ direction and polarized along the $z$ direction. The setting is as follows,

```
% plane wave
```

```
theta = linspace( 90, 90, 1);

phi = linspace( 0, 0, 1);

pol = 'TM';

% p excitation

p = multiexc( p, sim, theta, phi, pol, 'pw' );
```

As the second stage (as a separate simulation run), for the dipole excitation, the position, number and moment of the dipole (molecule) is defined in `fluoenhancement.m` and `quantum.m` (two independent subroutines called in `example1.m` for the calculation of normalized excitation rate and quantum yield). Here, the dipole is set at (0 nm, 0 nm, 90 nm) with a unit dipole moment along the *z* direction. Since the calculated results are normalized, the magnitude of the dipole moment does not affect the reported results.

```
% dipole

% pos ( positions and number of dipoles )

rd = 90 .* sim.nm;

msh.pos = [  0, 0, rd ];

msh.n = 1;

% pol ( dipole moment )

pd = [ 0, 0, 1 ];
```

For the calculation of the normalized excitation rate, the electric field at the dipole position is needed, which is composed of direct incident plane wave and scattered fields from the NP under the plane wave excitation. The parameters related to the interfaces and simulation are fed into the following subroutines, where solutions are recalculated under the dipole excitation.

```
% normalized excitation rate

eta = fluoenhancement( p, sim );

% quantum yield & Purcell Factor

[quany, purcellS] = quantum(p, sim);

% fluorescence enhancemenet

yemS = quany .* eta;
```

After all the initial setup, upon execution, the normalized excitation rate, quantum yield and Purcell factor are reflected by parameters `eta`, `quany` and `purcellS`, respectively through `fluoenhancement.m` and `quantum.m`, while the fluorescence enhancement `yemS` can be obtained by simple multiplication. By comparison, the results are also calculated by using a reference method, to be specific, the cascaded **S** matrix via the Redheffer Star Product (RSP) [43], through `fluorescenceenhancement.m`. The contrast is shown in Fig. 5. A good agreement can be observed, with the maximum relative error of the fluorescence enhancement rate equal to 0.07042 %, the maximum relative error of quantum yield equal to 0.07036 % and the maximum relative error of Purcell factor equal to 0.0870 %. The relative error is calculated as the value difference of the semi-analytical and the reference RSP solution, normalized by the value of analytical results.

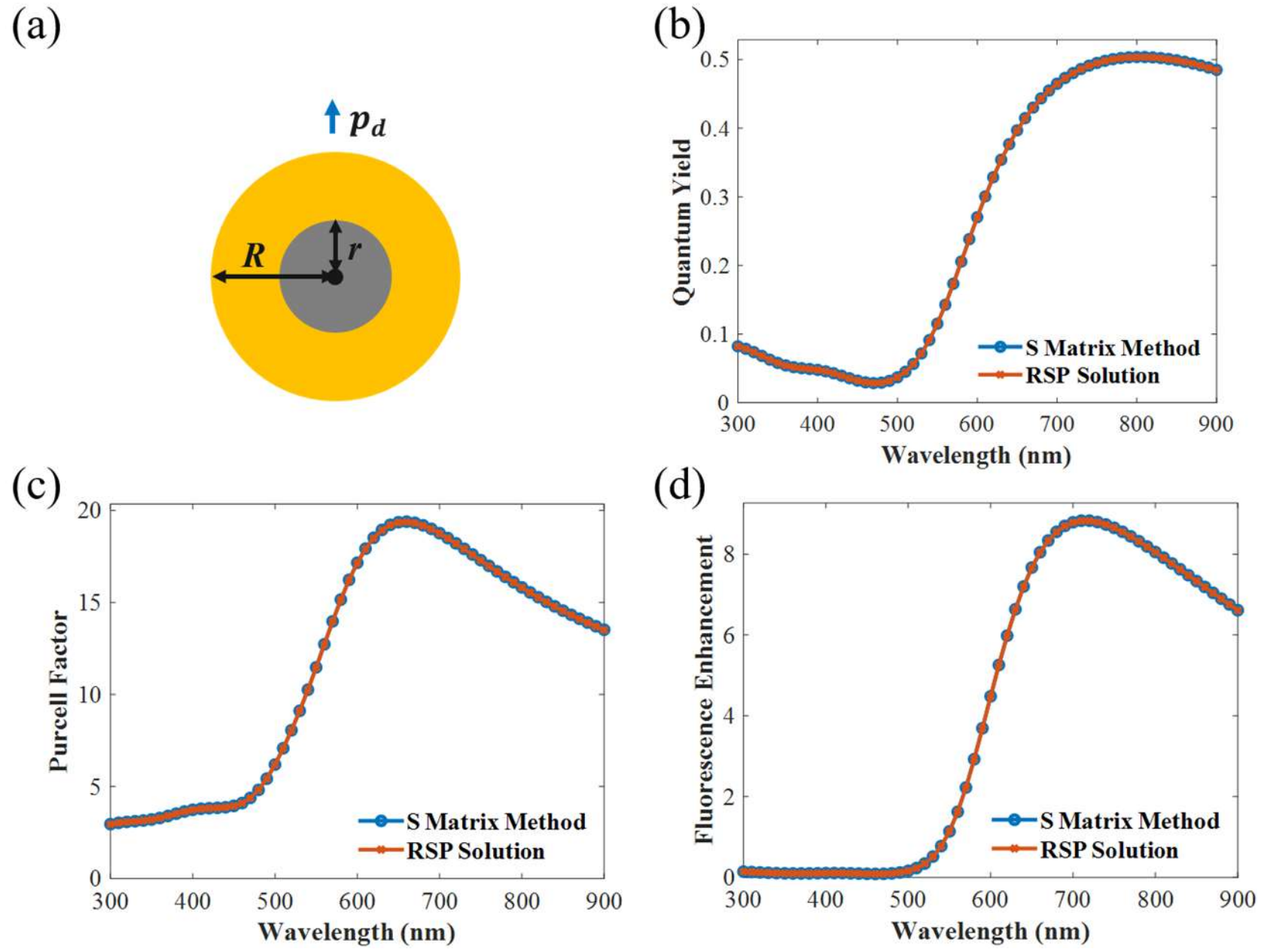

Fig. 5. (a) Illustration of the concentric nanosphere. (b-d) Comparison of the calculated quantum yield, Purcell factor and fluorescence enhancement calculated by the proposed toolbox (simi-analytical, blue line) and via the RSP (red line).

## 4.2 CL and EEL

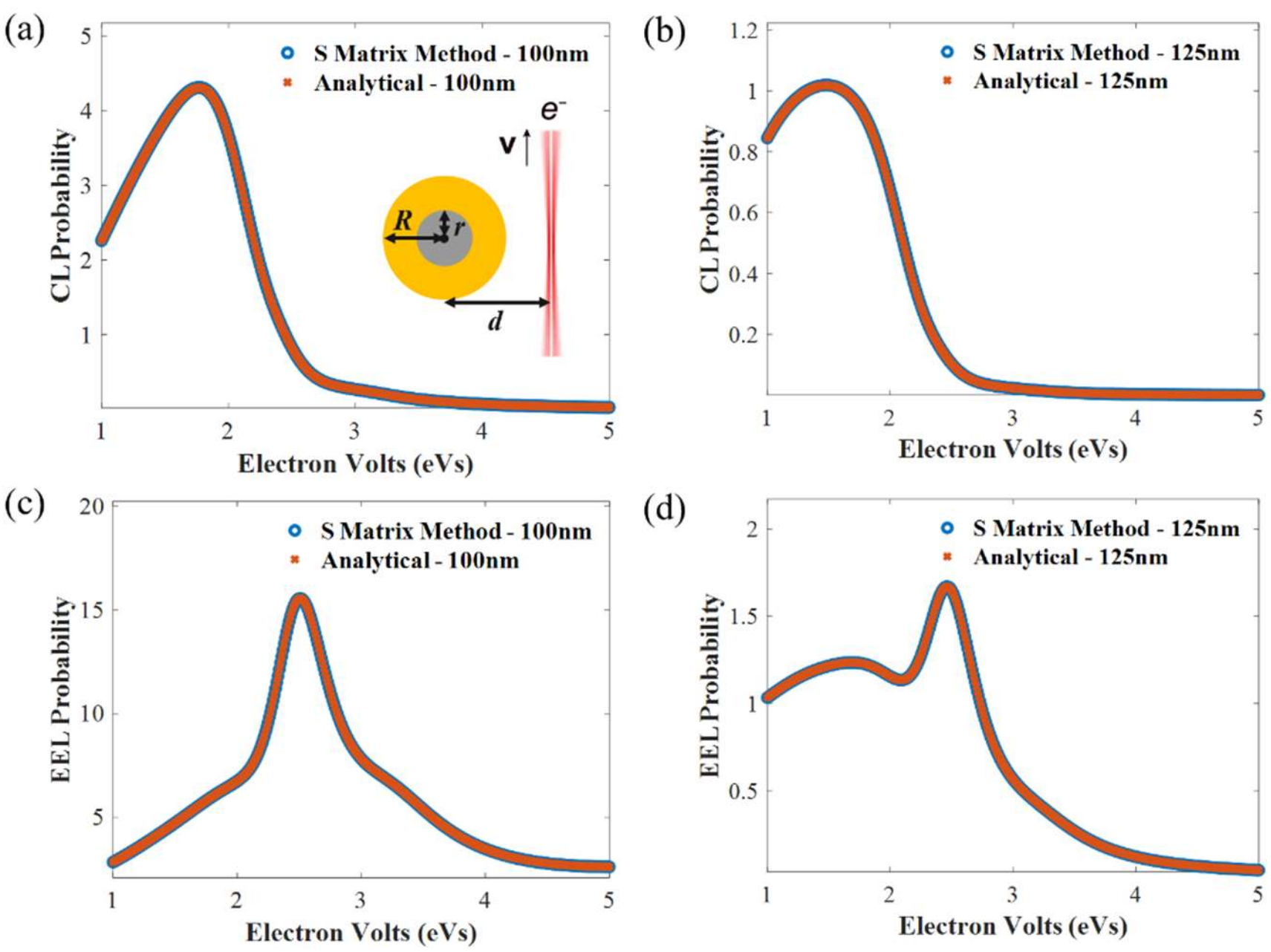

Fig. 6. Comparison of the calculated CL and EEL probabilities calculated by the proposed S Matrix toolbox (simi-analytical, blue circles) and analytical results (red crosses), for the concentric nanosphere shown in the inset of (a).

(a)(c) Results of the electron beam excitation at the position (100 nm, 0 nm, 0 nm). (b)(d) Results of the electron beam excitation at the position (125 nm, 0 nm, 0 nm).

In the second example, the same structure is excited by electron beams. The set-up of the simulation can be found in `example2.m` provided in the source-code package. To use this toolbox with electron beam excitation, the users need to define the simulated frequency band, the geometry, as well as the trajectories and speed of the electron beams. Once again, we must emphasize that we assume 1) the electron beam is incident beside the structure; 2) the electrons move along the $z$ axis. First, for the setup, an energy window ranging from 1 eV to 5 eV with 401 points taken in between is considered and the maximum harmonic order is 10.

```
% enei ( electron volts )
enei = linspace( 1, 5, 401 );
% lmax
lmax = 10;
```

Here, a concentric nanosphere in vacuum is discussed. The core is made from a dielectric material with a permittivity equal to 2.8, while the shell is made from Au, described by Lorentz-Drude Model, shown in the inset of Fig. 6(a). The setting of the geometrical structure is the same as the last example. The radius of the dielectric core `r1` is 25 nm, and the radius of the whole nanosphere `r2` is 85 nm. The center of the concentric nanosphere is set at origin (0 nm, 0 nm, 0 nm).

```
% bulk
sim.bulk = { materialconst(        2.8,  sim ); ...
             materialnlocal( 'Au', 'LD', sim ); ...
             materialconst(          1, sim ) };
```

For the excitation, two electron beams are used to excite the nanostructure, at the position of (100 nm, 0 nm, 0 nm) and (125 nm, 0 nm, 0 nm) at one third light speed. The electron velocity `v` is a key physical input parameter that determines phase matching, field decay, and excitation efficiency in electron–nanoparticle interactions. The last input term (which controls the "excitation mode") in `multiexc.m` needs to be set to `'eels'`,

```
% r
r = [100, 0, 0; ...
   125, 0, 0 ];
% speed
v = [0.33; 0.33] * sim.c;
% p
p = multiexc( p, sim, r, v, 'eels' );
```

After all the initial setup, upon execution, the calculated CL probability is stored in parameters `CL0`, through the `cl` function.

```
%% CL
```

```
% CL S Matrix Method

CL0 = cl( ff, p, sim);

% CL Analytical

CL1 = p.mie.cl ./ sim.ev;
```

The calculated EEL probability can be found in `p.eels`.

```
%% EELS

% scattering

p = eels(p, r, sim);

% EELS

EELS0 = p.eels / sim.ev;

% EELS Analytical

EELS1 = p.mie.eel.eel ./ sim.ev;
```

In the above, the CL probability and EEL probability are also calculated by using an analytical method [54], through `mieconcentricsphere.m`, stored in `CL1` and `EELS1`, respectively. The comparison of the analytical results and the results obtained by using the proposed MATLAB toolbox is shown in Fig. 6. For the electron beam at (100 nm, 0 nm, 0 nm), the maximum relative error of CL and EEL probability is 0.08750% and 0.00022%, respectively. For the electron beam at (125 nm, 0 nm, 0 nm), the maximum relative error of CL and EEL probability is 0.08755% and 0.00024%, respectively.

### 4.3 Circular Dichroism

In the third example, we check from a physical perspective to illustrate CD in a nanohelix. The polarization of plane waves is set as left-circular polarized or right-circular polarized, incident along the *z* axis. Users need to define the simulation frequency band, the material parameters, as well as the geometry. The simulation set-up can be found in the following four `m` files: `example3_LCP_gap03nm.m`, `example3_RCP_gap03nm.m`, `example3_LCP_gap05nm.m and example3_RCP_gap05nm.m`.

First, as inspired by [59], we focus on a wavelength range from 300 nm to 500 nm with 101 points taken in between and the maximum harmonic order is 22 to ensure the convergence,

```
% enei ( wavelengths )

enei = linspace( 300, 500, 101 );

% lmax

lmax = 22;
```

The NPs are made from silver, described by a dielectric function whose imaginary part is artificially reduced by 90%. The detailed material properties are described by `materialtab.m`, while the material parameters are obtained from the experimental dielectric function of Johnson and Christy [83], indicated in `silverlumerical.m`. Note that similar to [59], the 90% reduction of the imaginary part of silver's permittivity is to suppress intrinsic damping channels, allowing highly overlapping higher-order modes and chiral dark states to be cleanly resolved in the spectra. The used materials are stored in the `sim.bulk` variable as below. In the `sim.bulk` variable, "`materialtab( 'silverlumerical', sim )`" is repeated nine times, as they are identified as different materials (that is, *medium 1, 2, 3, 4, 5, 6, 7, 8 and 9*) by the solver.

```
% bulk

sim.bulk = { materialtab( 'silverlumerical', sim ); ...

             materialtab( 'silverlumerical', sim ); ...

             materialtab( 'silverlumerical', sim ); ...

             materialtab( 'silverlumerical', sim ); ...

             materialtab( 'silverlumerical', sim ); ...

             materialtab( 'silverlumerical', sim ); ...

             materialtab( 'silverlumerical', sim ); ...

             materialtab( 'silverlumerical', sim ); ...

             materialtab( 'silverlumerical', sim ); ...

             materialconst(                1, sim ) };
```

There are in total nine spheres, and the spheres revolve around the $z$ axis by a $\pi/2$ step, so that 9 NPs produce two full rotations, as shown schematically in Fig. 7(a). The rotation of the center of spheres due to rotation is described by `Rot.m` which is complemented by a translational displacement along the $z$ axis. The structure suggests a supporting pillar of diameter $d$ (see Fig. 7(a)), which is not included in the simulations. This type of structure can be made by using the DNA-origami technique [84]. The selected NP size fulfills two roles: when using nonlocal response models, it is sufficiently small for quantum effects to remain significant even without NP interactions, yet large enough to be compatible with standard DNA-origami pillars. The center of each NP is vertically shifted by $R$ (the radius of nanospheres, in this demo, it is 5 nm), while $d$ is 2.73499 nm, leading to a NP distance $g$ of 0.3 nm. We also simulate the situation when $g$ is equal to 0.5 nm, $d$ is 3.05756 nm. The corresponding setup in the simulation is shown below,

```
%% helix

% nn ( number of particles )

nn = 9;

% d, R

d = 2.73499; R = 5;

% set up pos

pos = zeros( nn, 3 );

% pos( 1, : ) ( the 1st sphere )

pos( 1, : ) = [ -( d/2 + R ), 0, 0 ];
```

```
% for the rest particles

for ii = 2 : nn

    % pos( ii, : )

    pos( ii, : ) = ( ii - 1 ) * [ 0, 0, R ] + ( Rot( ( ii - 1 ) * pi/2 ) *
pos( 1, : ).' ).';

end
```

Then, for each sphere, the geometry and the constituent materials are set up. Here, the code for the first nanosphere is given as an example, which can be done in a similar way for the other spheres.

```
% particle 1 ( center, radius, in/out materials, on-boundary material )

rc1 = pos( 1, : ); r1 = 5; inout1 = [ 1, 10 ]; on1 = 1;

% p1

p1 = particle( rc1, r1, inout1, on1, sim );
```

The system is illuminated by a circularly polarized plane wave propagating along the helix axis (*z* axis, accordingly `theta = 0` *and* `phi = 0`). The polarization is either right-circular polarized (that is, `'RCP'` in the simulation) or left-circular polarized (`'LCP'` in the simulation),

```
% theta

theta = linspace( 0, 0, 1 );

% phi

phi = linspace( 0, 0, 1 );

% pol

pol = 'RCP';

% p

p = multiexc( p, sim, theta, phi, pol, 'pw' );
```

Then, the calculated scattering cross sections and absorption cross sections are stored in parameters `sca0` and `abt0`, through `ffmapping.m` and `absorption.m`, respectively.

```
%% cross sections

% scattering

p = ffmapping( p, sim );

% absorption

p = absorption( p, sim );

% sca0

sca0 = p.ff.sca * 1e18;

% abt0
```

```
abt0 = p.nf.abt * 1e18;
```

After that, we calculate the CD as the difference of the absorption cross sections of LCP and RCP, normalized by $S(\pi R^2)$.

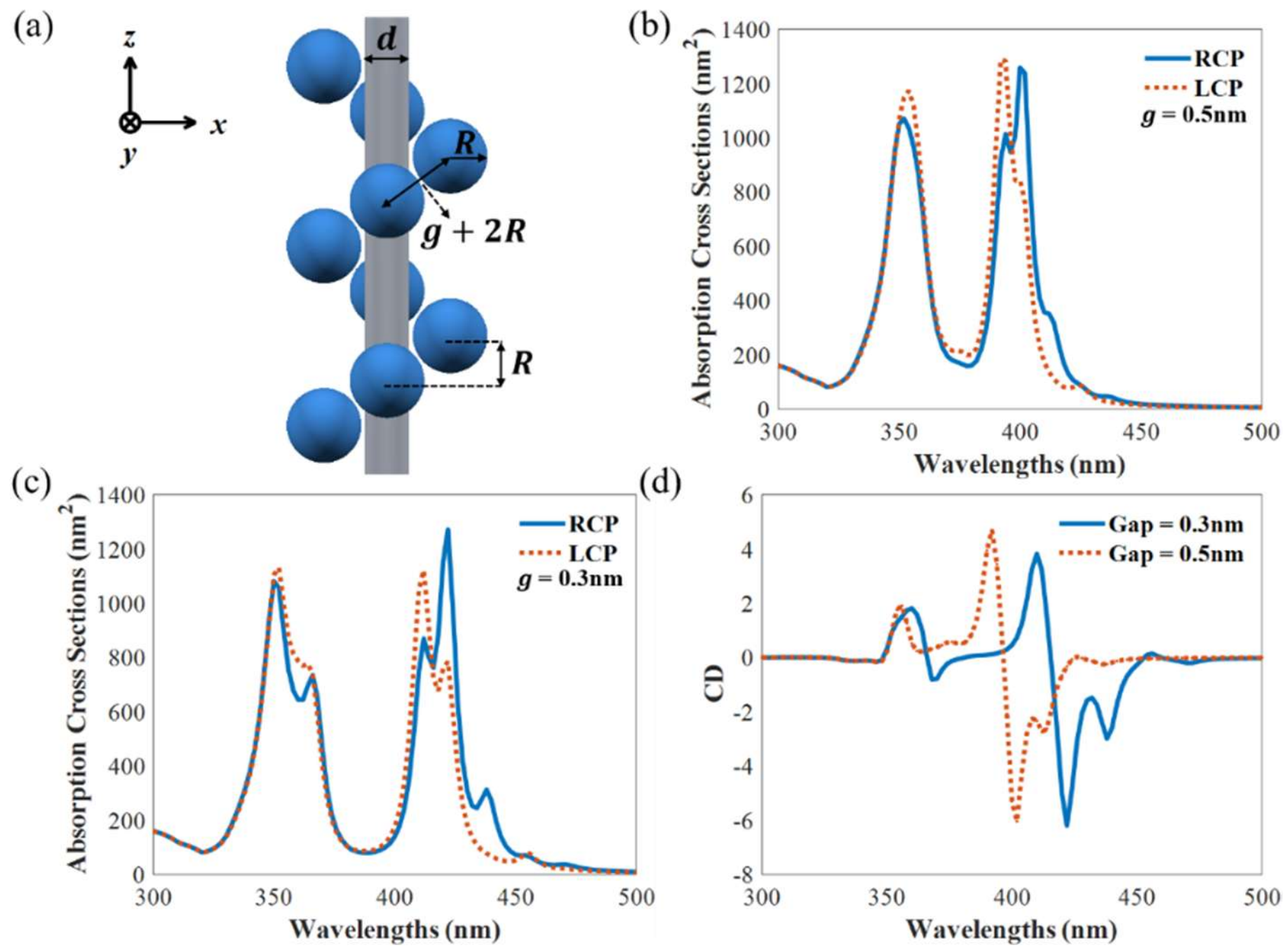

Fig. 7. (a) The helix structure. (b) Contrast of the calculated absorption cross sections under the excitation RCP plane wave against the excitation of LCP plane wave, when the gap between nanospheres is equal to 0.5 nm. (c) The contrast when the gap between nanospheres is equal to 0.3 nm. (d) Calculated CD for gaps being 0.3 nm and 0.5 nm.

The calculated results are shown in Fig. 7 and are in accordance with previous findings [59]. The resonance in the short wavelength is associated with the localized surface plasmon resonance of a single NP, while the strong interaction from 380 nm to 410 nm in Fig. 7 (b), and from 400 nm to 430 nm in Fig. 7 (c) resulting from the narrow gaps, can be understood in view of the embedded-chain model as collective chain modes in the direction of the electric field of the incident plane wave [85]. Here we define the LCP unit vector as $(\hat{x}+i\hat{y})/\sqrt{2}$, and the RCP unit vector as $(\hat{x}-i\hat{y})/\sqrt{2}$. When the gap is 0.5 nm, a double peak appears at 392 nm and 400 nm, while for the case when the gap is 0.3 nm, the double peak appears at 412 nm and 422 nm. The spectral split resulted from different polarization illuminations on the NP helix, which lacks mirror symmetry, leading to the CD signal shown in Fig. 7 (d). Though a frequency shift due to the change of gap size exists, the transition from peaks to dips in the CD spectrum remains obvious. Note that a large increase of the runtime for the helix structure is primarily due to 90% reduction of intrinsic Ohmic loss, which significantly increases the matrix condition number and requires more GMRES iterations per wavelength to achieve a strict convergence tolerance of $10^{-1}$ .

Since the gap is very small, we also perform a check on the convergence of the proposed toolbox. When $l_{max}$ increases from 22 to 24, the calculated cross sections stay unchanged, while the results obtained by lower $l_{max}$ deviated, as shown in Fig. 8, proving that the solver is highly converged. This physical check can be reproduced by executing the four demos, which leads to the spectral data shown in Fig. 7.

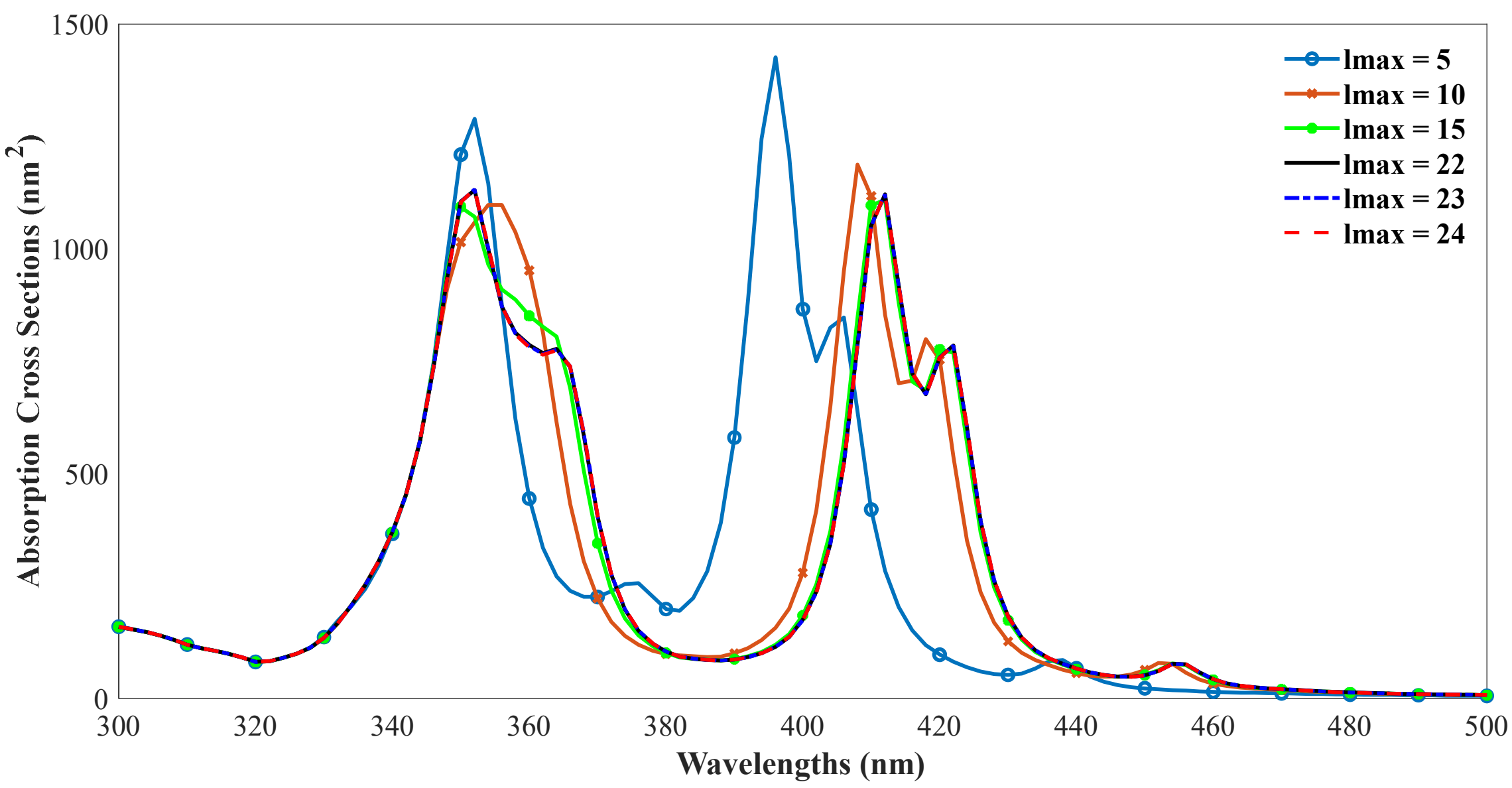

Fig. 8. Calculated absorption cross sections obtained by different maximum harmonic orders, proving the convergence of the obtained results.

## 5 Summary and Outlook

To summarize, this paper presents a MATLAB toolbox, suited for the simulation of multiple nanospheres with arbitrary centers (not partly overlapping), using the $\mathbf{S}$-Matrix Method and translational addition theorem. The toolbox combines three physical models (LRA, HDM and SRM) and three common excitation methods (plane waves, dipoles and electron beams) in the optical regime. Three examples are given, to calculate the fluorescence enhancement, Purcell factor and quantum yield of a molecular emitter (modeled as a dipole) near a concentric nanosphere under the excitation of plane waves; to calculate CL and EEL probability of a concentric nanosphere under the excitation of electron beams; and to calculate cross sections, as well as CD, of a helix structure under the excitation of LCP or RCP plane waves. The toolbox provides a guideline for the simulation setup. Users can easily adapt these demo files for other applications. Since nonlocal effects are also explored in this toolbox, the proposed toolbox serves as a perfect platform for the discovery of typical quantum effects in the nanoplasmonics field, and the calculated results can serve as a reliable reference for theoretical analyses or experimental verification.

For the nanosphere aggregates, regarding simulation time and accuracy, we believe that the toolbox performs well and can compete with other simulation solvers used in the plasmonic community. More nonlocal response models can be incorporated, for example, 1) GNOR, to include the diffusive effects, by modifying the longitudinal material parameters (adding a diffusion related term [12] to the expression of the longitudinal wavenumber and dielectric function), requiring correspondingly adapted boundary conditions; 2) the quantum hydrodynamic Drude model, to include the spill-out effect, by extracting the corresponding $d$ parameters [30,86]. In addition, the simulation of more complicated topologies, such as non-spherical scatterers, can be further explored based on the null-field method [87]. With its present framework and the prospective enhancements discussed above, we envision this toolbox serving as a powerful asset for researchers in nanophotonics, facilitating the exploration of exotic and extreme nano-optical effects.

# Appendix

## A. The pre-generation and storage of a set of translation coefficient tables

The pre-generation and storage of the translation coefficient tables are finished in the `init_coeffs.m`. First, we set up an upper limit of the maximum order `lmax` and the number of division groups `n0`.

```
% lmax

lmax = 29;

% n0

n0 = 3;
```

Then, all spherical harmonic modes $(l, m)$ up to the maximum order (29 here, and can be adjusted to larger numbers if needed) are divided into `n0` groups, each covering a continuous range of 10 $l$ values. After the grouping is completed, we consider every pair of groups $(jj, ii)$, where $ii, jj = 1, 2, \ldots, n_0$. For each pair, we take all $(l_1, m_1)$ modes from group $ii$ and all $(l_2, m_2)$ modes from the group $jj$ and compute the corresponding translation coefficients for every possible combination.

The actual coefficient calculation is performed by the `Tcoeff` function, and the outputs are summarized in Table A1. Each block contains the corresponding $(l_3, m_3)$ and numerical coefficients used for constructing the translation matrix. All coefficients for a group pair $(jj, ii)$ are saved into a separate `Tcoeffs_jj_ii.mat` file, forming a block-structured database of translation tables. An additional summary file `Tcoeffs.mat` records the range of $l$ of each group. During simulation, the program can quickly load the blocks it needs, instead of recomputing everything from scratch.

Table A1. Translation Coefficient Blocks.

| | $(l_1, m_1)$ Section 1 (0 - 9) | $(l_1, m_1)$ Section 2 (10 - 19) | $(l_1, m_1)$ Section 3 (20 - 29) |
|---|---|---|---|
| $(l_2, m_2)$ Section 1 (0 - 9) | `Tcoeffs_1_1` | `Tcoeffs_1_2` | `Tcoeffs_1_3` |
| $(l_2, m_2)$ Section 2 (10 - 19) | `Tcoeffs_2_1` | `Tcoeffs_2_2` | `Tcoeffs_2_3` |
| $(l_2, m_2)$ Section 3 (20 - 29) | `Tcoeffs_3_1` | `Tcoeffs_3_2` | `Tcoeffs_3_3` |

## B. Scaling Study

To evaluate both the computational efficiency and numerical robustness of the newly implemented mesoscopic framework, this appendix presents a systematic scaling study regarding the SRM, where two tests with respect to $l_{\max}$ and the number of spheres $N$ are performed. These simulations can be easily performed with a few minor modifications to the inputs, using the codes of example 3 as the template. In the first example, a sodium dimer structure with a radius of 10 nm is modeled using SRM, and the gap is equal to 1 nm. The dimer structure is excited by TM polarized plane waves. The number of $l_{\max}$ is increased to test the convergence of the simulation. The wavelength ranges from 400 nm to 800 nm, with 41 samples taken in between. This wavelength range setup is used in the rest of this section. Here, we list the runtime and memory scale used in the simulation.

Table B1. The scaling study with respect to $l_{\max}$.

| $l_{max}$ | Runtime (without post-processing)/s | Peak Memory Usage/MB |
| --- | --- | --- |
| 5 | 3.773 | 11.94 |
| 10 | 14.502 | 23.50 |
| 19 | 92.504 | 155.78 |
| 20 | 148.135 | 428.32 |
| 21 | 171.558 | 443.19 |

As $l_{\max}$ increases from 19 to 21, the calculated absorption cross section stays unchanged, while for lower orders, the results depart, as shown in Fig. B1, proving that this simulation is highly converged.

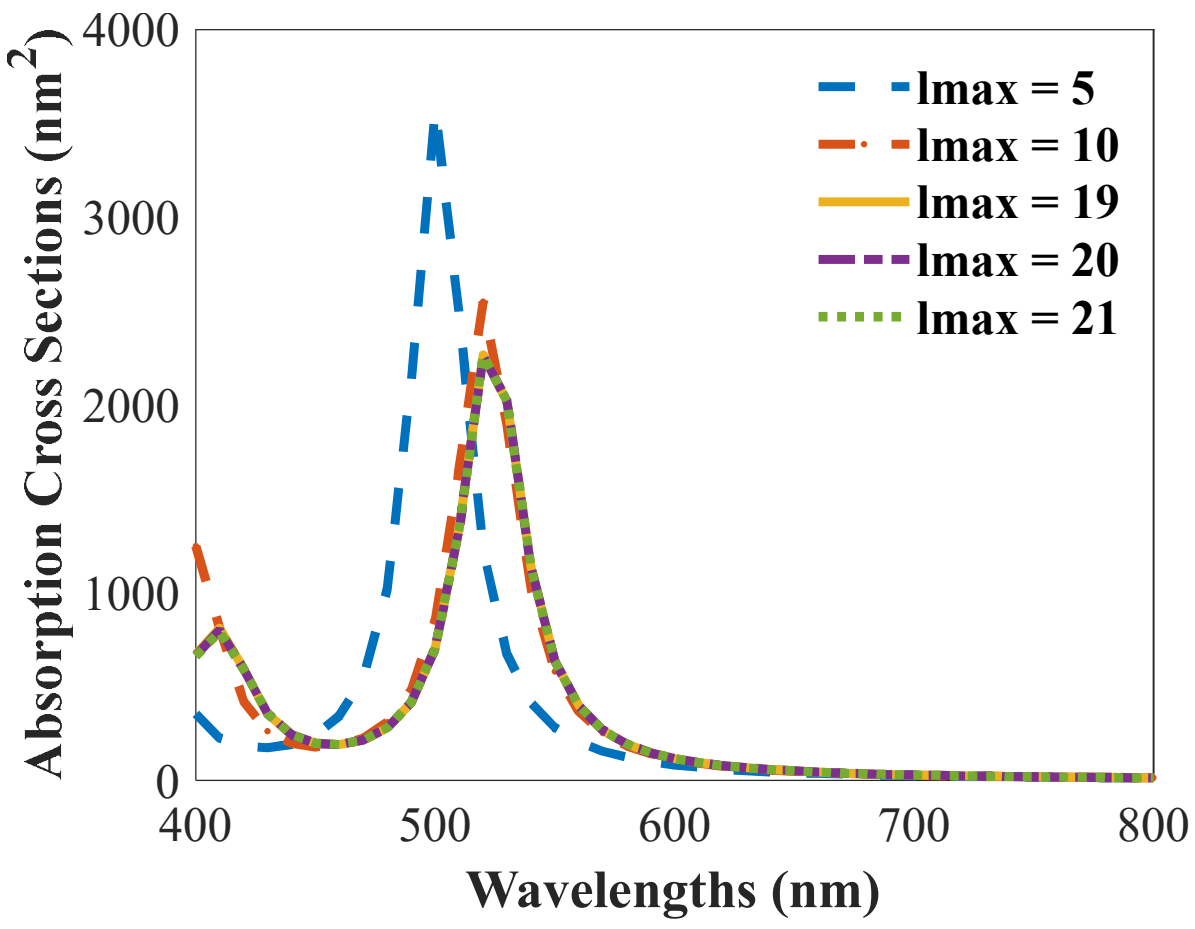

Fig. B1. The calculated absorption cross sections obtained by different $l_{\max}$, proving the convergence of the simulation.

In the second example, shown in Table B2, Na nanospheres with a radius of 10 nm are modeled. When additional spheres are introduced, their positions are arranged such that the gap between adjacent spheres is consistently maintained at 1 nm. Considering the complicated sodium trimer, the $l_{max}$ is set to 25 to guarantee the convergence of each test.

Table B2. The scaling study with respect to number of nanospheres.

| Number N | Runtime (without post-processing)/s | Peak Memory Usage/MB |
| --- | --- | --- |
| 1 | 98.701 | 374.22 |

| 2 | 347.136 | 445.23 |
|---|---|---|
| 3 | 863.332 | 1003.11 |

## Acknowledgement

Xin. Zheng is grateful for the China Scholarship Council (Grant No. 202206090027), China. Xuezhi Zheng, and Guy A. E. Vandenbosch are grateful for the C1 project (C14/19/083), the IDN project (IDN/20/014), the small infrastructure grant (KA/20/019) from KU Leuven and for G090017N and G088822N from the Research Foundation of Flanders (FWO). Xuezhi Zheng would like to also thank the support from the IEEE Antennas and Propagation Society Postdoctoral Fellowship, and for the FWO long travel grant, FWO V408823N. The Center for Polariton-driven Light – Matter Interactions (POLIMA) is funded by the Danish National Research Foundation (Project No. DNRF165) (Corresponding author: Xuezhi Zheng).

X. Zheng (e-mail: xin.zheng@esat.kuleuven.be) and G. A. E. Vandenbosch (e-mail: guy.vandenbosch@esat.kuleuven.be) are affiliated with the WaveCore Division, Department of Electrical Engineering (ESAT), KU Leuven, B-3001, Leuven, Belgium.

X. Zheng (e-mail: xuezhi.zheng@esat.kuleuven.be) is affiliated with Key Laboratory of Radar Imaging and Microwave Photonics of Ministry of Education, Nanjing University of Aeronautics and Astronautics, Nanjing, 211106, Jiangsu, China.

C. Mystilidis (e-mail: chrmys@mci.sdu.dk) and C. Tserkezis (e-mail: ct@mci.sdu.dk) are affiliated with POLIMA – Center for Polariton-driven Light-Matter Interactions, University of Southern Denmark, 5230 Odense, Denmark.